\documentclass[10pt,conference]{IEEEtran}
\IEEEoverridecommandlockouts
\usepackage[T1]{fontenc}
\usepackage[utf8]{inputenc}
\usepackage{cite}
\usepackage{amsmath,amssymb,amsfonts}
\usepackage{graphicx}
\usepackage{textcomp}
\usepackage{xcolor}
\usepackage{array}
\usepackage{booktabs}
\usepackage{tabularx}
\usepackage[hyphens]{url}
\usepackage[hidelinks]{hyperref}
\usepackage{xurl}

\title{AIPC: Evidence-Guided Agent Workflows for PyTorch Project Deployment}
\author{\IEEEauthorblockN{\parbox{\textwidth}{\centering Zhanwei Wu$^{1,*,\dagger}$, Jianhao Su$^{1,*}$, Brian Zhang$^{1,*}$, ShengTing Huang$^{1}$, Weidong Feng$^{1}$, Gang Liu$^{1}$, Zhiguo Li$^{1}$, Wei Huang$^{1}$, Li Tan$^{1}$, Wanjun Li$^{1}$, Zejiong Teng$^{1}$, Changgeng Zhong$^{1}$, Guanping Wu$^{1}$, Han-Hsing Tu$^{2}$}\thanks{$^{*}$These authors contributed equally to this work.}\thanks{$^{\dagger}$Corresponding author: Zhanwei Wu (zhanweiw@qti.qualcomm.com).}}
\IEEEauthorblockA{\parbox{\textwidth}{\centering $^{1}$Qualcomm Technologies, Inc.\quad $^{2}$Independent Researcher}}}
\begin{document}
\maketitle
\begin{abstract}
Deploying and optimizing a PyTorch inference project on a dedicated accelerator, especially an on-device NPU, usually cannot be done through the framework's device API alone as with CUDA; the project must also be converted into hardware-runtime artifacts and adapted to the target toolchain and application interface. Platforms differ in model formats, quantization flows, operator support, and toolchain interfaces, so model adaptation, numerical correctness, and hardware performance problems often have to be located, repaired, and rechecked back and forth across the model (PyTorch/ONNX), toolchain, runtime, and device-execution layers. LLM Agents can invoke tools to handle such cross-layer faults, but without project constraints they may still drift off task because of insufficient domain knowledge, an overly large space of configuration and repair choices, or poor path selection.

To this end, we propose \textbf{AIPC} (AI Porting \& Conversion), an Agent method for deploying PyTorch inference projects. AIPC combines structured Skills with toolchain, runtime, and profiling evidence to drive cross-layer diagnosis and repair, and organizes model adaptation, runtime artifact building, on-target execution, accuracy repair, and performance optimization into one deployment flow. We evaluate AIPC on the on-device NPU path of Qualcomm QAIRT, covering vision, speech, LLM, and diffusion models. In two optimization cases, AIPC's optimization knowledge and strategies guide the Agent to accuracy and performance gains: under the predefined accuracy and latency criteria, 6/7 Skill arm Whisper-Small runs meet all criteria versus 0/6 in the control arm; YOLO-Worldv2 reaches an end-to-end speedup of 2.791$\times$ (control arm 1.599$\times$). Under a pre-fixed deployment contract, AIPC Skill, tool permissions, and check conditions, some runs achieved zero in-run human intervention (0-HITL). We also run an auxiliary experiment to examine the ability of Qwen3.8-27B, a model capable of local deployment, to drive an Agent workflow.
\end{abstract}

\section{Introduction}\label{sec:1}

\subsection{Motivation and Engineering Pain Points}\label{sec:1.1}

On-device and edge AI inference provides AI PCs, smartphones, intelligent vehicles, industrial robots, and edge IoT devices with AI compute that offers low latency and high energy efficiency and supports offline operation and data locality. For device backends such as CUDA~\cite{r1} that are integrated into PyTorch~\cite{r2}, developers can execute models directly through framework interfaces; dedicated accelerators, especially on-device NPUs, usually require converting the model into hardware-runtime artifacts. However, model deployment spans frameworks, graph representations, conversion tools, runtimes, and device backends, whose interfaces, operator coverage, layouts, and inspection methods are not consistent. Deployment knowledge and failure information are therefore scattered among the model, compilation, runtime, and on-device execution layers, and as the numbers of models and target configurations grow, the expert cost of maintaining the deployment flow grows as well. This paper takes Qualcomm QAIRT as its primary implementation and evaluation target and examines the following coupled engineering problems:

\textbf{Challenge 1: Deployability --- ``Can it run?''}

On-device deployment requires converting the dynamic-graph semantics of PyTorch into artifacts accepted by the target runtime. Dynamic shapes, control flow, complex indexing, and non-standard operators may all cause conversion, compilation, or loading failures.

\textbf{Challenge 2: Accuracy Correctness --- ``Does it still compute the same function?''}

Successful conversion and build do not guarantee that the original model semantics are preserved. Problems such as all-zero outputs, NaN/Inf, tensor-order mismatches, and numerical drift after quantization must be detected through automated alignment checks against reference outputs.

\textbf{Challenge 3: Performance Optimization --- ``Does it really use the accelerator efficiently?''}

Running successfully does not mean running efficiently. Unsupported operators, layout conversion, and limited graph fusion cause CPU fallback, cross-device data movement, or extra memory access. Exploiting the target accelerator's performance also requires hardware-aware optimization.

These challenges force developers to iterate repeatedly among the model, conversion configuration, runtime build, and on-device checks, limiting deployment efficiency and scalability. AgentDeploy~\cite{r3} reports that experts spent around 6 h on image-recognition tuning alone, while semi-experts and non-experts required about 60 h and 80 h, respectively, for speech-sentiment-analysis optimization and adaptation. Agents can assist diagnosis and repair in this repeated cross-layer work, while project checks constrain execution.

\subsection{Limitations of Existing Solutions}\label{sec:1.2}

In response to these challenges, existing tools provide model conversion, compilation optimization, and code repair, but three key gaps remain among them:

\textbf{(1) Toolchain feedback does not form a repair loop (missing loop from tool feedback to repair)}: Converters, compilers, and runtimes can usually report on-device errors, numerical deviations, or backend information, but this information does not turn itself into repair actions. Developers still have to decide whether to modify the PyTorch model, the ONNX graph, or the runtime configuration, and then rebuild and re-verify. What is missing is the closed loop of ``toolchain feedback $\rightarrow$ cross-layer diagnosis $\rightarrow$ repair $\rightarrow$ rebuild $\rightarrow$ re-verification.''

\textbf{(2) Missing automated result verification}: Successful conversion and compilation do not guarantee correct output. Without automatically running inference, aligning with reference outputs, and checking the target backend, silent failures such as NaN/Inf, all-zero outputs, and CPU fallback are hard to detect systematically.

\textbf{(3) Missing deterministic execution boundary for open-ended diagnosis}: An LLM Agent (e.g., ReAct~\cite{r4}) can handle errors that were not enumerated in advance, but may also retry blindly or drift off task.

\subsection{AIPC Method Overview and Contributions}\label{sec:1.3}

Building on earlier AIPC reports~\cite{r5,r6}, this paper extends \textbf{AIPC (AI Porting \& Conversion)}, a runtime-feedback-driven Agent method for on-device model porting. AIPC organizes platform knowledge, execution scripts, stage checks, and deployment requirements into reusable Skills, and chooses the repair layer and subsequent path based on model, device, and runtime feedback.

The method of this paper includes:

\begin{enumerate}
\item \textbf{Feedback-guided deployment workflow for existing inference projects}: Taking a PyTorch project as the entry point, it brings the model, pre/post-processing, I/O contract, and target-device requirements into a single deployment process, and outputs runtime artifacts, project configuration, and run records.
\item \textbf{Feedback-guided cross-layer diagnosis, repair, and recheck}: Toolchain, runtime, numerical, and profiling feedback helps the Agent decide at which layer to repair --- the model graph, the project interface, or the runtime configuration --- and then rebuild and recheck the affected paths. For example, when AOT (ahead-of-time) compilation runs out of on-chip scratchpad memory, the Agent can localize the cause to the ONNX decoder graph. Deployment results are confirmed by project checks and run records.
\item \textbf{QAIRT proof-of-concept and reusable engineering lessons}: AIPC is implemented on the QAIRT on-target NPU, and representative cases report runtime-artifact build, on-target execution, output alignment, profiling, and failure recovery.
\end{enumerate}

This paper evaluates AIPC through the following research questions:

\begin{itemize}
\item \textbf{RQ1:} Can an Agent workflow complete target-device deployment of PyTorch projects? How does the AIPC Skill affect deployment paths and delivery outcomes?
\item \textbf{RQ2:} In representative execution-failure cases, how do runtime, toolchain, and output-comparison results trigger the Agent's cross-layer diagnosis, repair, and recheck?
\item \textbf{RQ3:} Can AIPC's accuracy and performance optimization knowledge and strategies guide the Agent to improve device-side performance while satisfying the accuracy requirements and keeping NPU execution?
\end{itemize}

\section{Background and Related Work}\label{sec:2}

\subsection{On-Device Conversion and Compiler Frameworks}\label{sec:2.1}

For accelerators integrated as PyTorch device backends, developers usually run models through interfaces such as \texttt{.\allowbreak{}to(device)} or \texttt{.\allowbreak{}cuda()}. Dedicated hardware runtimes usually do not execute models through these device APIs; instead, the project is exported as a computation graph, a runtime artifact is built, and inference runs through a hardware-specific runtime interface. This path helps exploit hardware performance, but it also introduces adaptation problems.

Frameworks such as ONNX~\cite{r7}, ONNX Runtime~\cite{r8}, ExecuTorch~\cite{r9}, LiteRT~\cite{r10}, and MNN~\cite{r11} provide conversion and inference infrastructure; TVM~\cite{r12}, MLIR~\cite{r13}, IREE~\cite{r14}, and Microsoft Olive~\cite{r15} provide graph, kernel, or configuration optimization. They typically do not turn device-specific conversion, runtime, numerical, or profiling feedback into corresponding repairs and rechecks.

Chip and platform vendors also provide dedicated inference runtimes for their devices, such as NVIDIA TensorRT~\cite{r16}, Qualcomm QAIRT~\cite{r17}, Apple Core ML~\cite{r18}, Rockchip RKNN~\cite{r19}, Intel OpenVINO~\cite{r20}, and Huawei Ascend CANN~\cite{r21}. These runtimes combine model conversion, hardware-specific compilation, and target-device execution, but their artifacts and interfaces differ.

\subsection{Qualcomm QAIRT Toolchain and Application Runtime}\label{sec:2.2}

Qualcomm AI Runtime (QAIRT)~\cite{r17} is Qualcomm's official inference software stack for heterogeneous SoCs that integrate a CPU, Adreno GPU, and Hexagon NPU, covering vision models through generative models; HTP (Hexagon Tensor Processor) is the tensor-compute accelerator of the Hexagon NPU, and QAIRT can dispatch supported models to this backend.

\textbf{Three main model artifacts.} (1) \textbf{DLC (Deep Learning Container, \texttt{.\allowbreak{}dlc})}: a serialized graph container; (2) \textbf{QNN model library (\texttt{.\allowbreak{}so}/\texttt{.\allowbreak{}dll})}: generated by target compilation of the C++ model source and static weights output by the converter, and therefore dependent on the compiler, ABI, operating system, and target architecture; (3) \textbf{Context Binary}: an execution context generated from a DLC or QNN model library by ahead-of-time (AOT) precompilation for a target backend/SoC and saved as a \texttt{.\allowbreak{}bin}.

\textbf{APIs and toolchains.} The API layer is responsible for loading and executing graphs, operators, and runtime resources. (1) \textbf{SNPE API}: typically uses a DLC and is handled by the \texttt{snpe-\allowbreak{}*} toolchain; (2) \textbf{QNN API}: can use a target model library or a DLC and is handled by the \texttt{qnn-\allowbreak{}*} toolchain; (3) \textbf{QAIRT API}: typically uses a DLC or a prepared Context Binary and is handled by the \texttt{qairt-\allowbreak{}*} toolchain.

\textbf{Genie (Generative AI Inference Extension)}~\cite{r17} is a dedicated inference extension for on-device LLMs, vision-language multimodal models, and diffusion models.

\textbf{AIMET and AI Hub.} AIMET (AI Model Efficiency Toolkit)~\cite{r22} supplements QAIRT's built-in quantization flow with finer-grained quantization, calibration, and model-compression capabilities; Qualcomm AI Hub~\cite{r23} is a cloud platform for model compilation and performance evaluation for Qualcomm devices.

\textbf{QAI AppBuilder, Model Builder, and AIPC.} QAI AppBuilder~\cite{r24} is an on-device AI application platform built on QAIRT and open-sourced by Qualcomm (BSD-3-Clause). It provides Python and C++ APIs on the target side, which can load QAIRT model artifacts, configure the target backend, and run inference, so that tensors, graphs, buffers, and layouts need not be managed directly through the lower-level C++ QAIRT SDK API. Model Builder is its built-in package for model conversion and deployment. AIPC is a Skill-based workflow method inside Model Builder for deployment work.

A typical deployment chain includes: (1) PyTorch-to-ONNX export; (2) conversion and runtime-artifact build; (3) optional quantization and Context Binary preparation; (4) application-interface integration and on-target inference; and (5) profiling-driven iterative optimization.

\textbf{Hardware and toolchain characteristics.}~\cite{r17} (1) The VTCM (Vector Tightly Coupled Memory) of HTP provides low-latency on-chip storage for compute tiles; when a single tile exceeds the available capacity, graph compilation may fail or off-chip data access may increase. (2) The QNN converter may change ONNX I/O to a Spatial-First layout (e.g., NHWC), thereby changing the I/O contract; the QAIRT converter usually preserves the source layout. Unsupported operators, dynamic shapes, or complex control flow may cause failures at the conversion, compilation, or execution stage.

Beyond the direct QAIRT SDK path, HTP is also available through the QNN integrations of ONNX Runtime, LiteRT, ExecuTorch, MNN, and llama.cpp~\cite{r25}, fragmenting deployment paths for the same hardware.

\subsection{Agentic Deployment and LLM Agent Related Work}\label{sec:2.3}

In recent years, LLM Agents have been used for NPU deployment, runtime integration, and edge inference. Prior work~\cite{r5} explored an Agent-driven, toolchain-centered QAIRT deployment workflow, and~\cite{r6} applied the workflow to vision models on OpenVINO, RKNN, TensorRT, and ONNX Runtime QNN EP. This paper further extends PyTorch project adaptation and adds accuracy-repair and performance-optimization capabilities.

Chip vendors have also begun to encode hardware and software-stack knowledge as Agents or Skills to support operator development, performance diagnosis, model migration, and inference optimization. AMD's ROCm AI provides reusable Agent Skills~\cite{r26}, and Hyperloom~\cite{r27} optimizes inference workloads on AMD GPUs; Huawei's Ascend/agent-skills~\cite{r28} and CANNBot-Skills cover model conversion, inference deployment, profiling, and Ascend C operator development; and Intel Xe-Forge~\cite{r29} targets kernel optimization on Intel XPUs.

Spatial-NPU Agent Skill~\cite{r30} is based on MLIR-AIR and uses a staged Agent Skill to support the deployment and optimization of decoder-only LLMs on the AMD XDNA 2 NPU; AgentDeploy~\cite{r3} mainly searches tunable optimization parameters such as the pruning rate and partitions models across multiple inference engines according to operator support; and EdgeFM~\cite{r31} targets edge VLM/LLM kernel optimization. Related work also starts from kernel, operator, or configuration search and optimizes using compilation, correctness, or profiling feedback~\cite{r9,r29,r32,r33,r34,r35,r36,r37,r38,r39,r40}. AIPC, in contrast, takes an existing PyTorch inference project as its unit of work and organizes cross-model-type adaptation, runtime model and code generation, on-target inference, accuracy repair, and performance optimization into a complete workflow.

\section{AIPC System Framework}\label{sec:3}

AIPC comprises two execution components: the \textbf{Model Builder} and the \textbf{Optimizer}. The Model Builder builds a PyTorch inference project into a deployment baseline on the target device; the Optimizer takes this baseline as input and generates optimized variants that satisfy accuracy and target-backend constraints. The Agent calls these components and, guided by the AIPC Skill, reads state, selects tools, and proposes repair candidates.

AIPC organizes a deployment workflow around \textbf{Config}, \textbf{Pass}, and \textbf{Gate}. Config fixes the project entry, I/O contract, target device, runtime paths, and resource constraints; a Pass is a configurable stage unit that defines inputs, outputs, checks, Gates, and candidate actions; a Gate is a status value derived from artifacts and run results that triggers a Pass to continue, repair, transfer to another Pass, or stop. Passes form the workflow; in project setup, the engineer can skip or add steps as the project requires, and the Agent selects tools and candidate repairs within this flow.

\subsection{Model Builder}\label{sec:3.1}

The Model Builder takes a PyTorch project, deployment Config, and Gate conditions as inputs; its outputs are the target runtime artifacts, the project interface adaptation, and the deployment baseline. It executes in the Pass order determined in setup and records Pass state and artifact dependencies, rather than treating export, conversion, and on-device execution as independent scripts. The target runtime defines its corresponding default Passes; a project may skip or add steps as needed (for example, the FP16 path skips quantization), but must preserve their required inputs and Gates.

The default Passes of the QAIRT path are: P0 project and environment configuration, P1 PyTorch/NPU adaptation, P2 ONNX export and reference preparation, P3 graph check and operator repair, P4 conversion, quantization, and runtime build, P5 Context Binary preparation, P6 application interface and on-target inference, P7 accuracy report, and P8 profiling. P1, P3, and P6 handle project-level work beyond SDK-tool conversion and runtime-artifact generation.

\textbf{P1: PyTorch/NPU adaptation.} Before export or SDK-converter invocation, P1 handles representation differences between the source project and the target NPU, for example fixing dynamic shapes or control flow, externalizing the autoregressive cache, establishing an exportable model entry, and, when needed, rewriting RoPE or Multi-Head Attention into static tensor computation better suited to execution on the target NPU.

\textbf{P3: Graph check and operator surgery.} P3 turns unsupported structures or operators reported by an exporter or converter into graph-repair tasks. The Agent can choose an equivalent rewrite whose preconditions are satisfied at the PyTorch, export-entry, or ONNX-graph layer; the concrete repair layer is determined by project configuration and current feedback. Each rewrite preserves the input/output conditions, and reference-output comparison, subsequent conversion, and target execution checks decide whether it proceeds to the next Pass.

\textbf{P6: Application interface and on-target inference.} P6 connects generated runtime artifacts, preprocessing, model invocation, and postprocessing to the on-target inference code, completing runtime model loading, I/O binding, and on-target execution. Specialized application APIs, such as the Genie API for LLMs, can also be handled by an additional API Pass.

When a Gate fails, the Model Builder routes the problem to the PyTorch, ONNX, or runtime-interface layer based on logs, output comparison, and runtime feedback, and re-runs from the affected Pass. Graph rewriting, operator replacement, graph partitioning, and I/O/layout repair all belong to the Builder's deployment repair. Each run retains the Config, Pass artifacts, command logs, and comparison results, so that later Passes can reuse existing state rather than having the Agent re-guess project conditions.

\subsection{Optimizer}\label{sec:3.2}

The Optimizer is an internal, non-open-source extension that the company's internal customer-support engineering team has used to provide model-optimization support to external customers. It takes a deployment artifact as its baseline and uses profiling, VTCM, operator behavior, numerical or task feedback, and the graph-optimization knowledge substrate to select quantization adjustments, ONNX equivalent graph surgery, or runtime-configuration candidates. The knowledge substrate encodes graph-rewrite patterns and their applicability preconditions, hardware-benefit hypotheses, historical positive and negative cases, and candidate-selection rules. A candidate becomes the new optimization baseline after rebuilding and passing the acceptance Gate; candidates that do not satisfy accuracy, project I/O, or target-NPU execution requirements are rejected.

\subsubsection{Accuracy Optimization}\label{sec:3.2.1}

Accuracy optimization uses the FP reference, target-device outputs, and numerical error or task metrics to locate overall deviation, sensitive outputs, and task-accuracy degradation, and requires A/B comparisons that change one variable at a time, with the activation and weight calibration methods treated as two independent variables. For problems that are hard to localize, the Optimizer can further analyze clipping effects, layer-wise output deviations, and residual outliers. Candidates include calibration data and methods, per-tensor/per-channel/per-row quantization granularity, global low precision or mixed precision, FP16 retention of sensitive layers or output heads, and equivalent graph rewrites for numerical repair.

\subsubsection{Performance Optimization}\label{sec:3.2.2}

Performance optimization locates bottlenecks based on profiling, VTCM usage, operator behavior, and data movement. Candidates include layout optimization (NCHW$\leftrightarrow$NHWC and transpose elimination), operator-level equivalent rewriting (Einsum decomposition, Mod rewriting, and Cast substitution), attention restructuring (MHA$\rightarrow$SHA), and operator fusion (GELU subgraph, BN folding, parallel-convolution and branch merging), as well as VTCM configuration or graph partitioning.

\subsection{Usage Modes}\label{sec:3.3}

The workflow starts from a runnable PyTorch inference project, reference inputs, an inference script, and model weights, and delivers a hardware-runtime script, target model artifacts, and a work report with check, accuracy, and performance records; task results can be confirmed by human review or an independent evaluation program.

Users start tasks in natural language. For project-level deployment, the user first states the target device, precision, or optimization requirements, for example ``Set up a deployment project using QNN FP16''; the Agent initializes the working directory and Config accordingly and generates candidates for the remaining derivable configuration. For parameters that cannot be derived, such as LLM context length, the Agent may first propose a candidate value, or stop on a blocking condition and request user input. After confirmation, the Agent can run continuously (continuous mode), or pause after each stage and continue only upon user confirmation (interactive mode); the user can also state the deployment goal and constraints in a single prompt. Single tasks (for example, generating a Context Binary for an audited quantized DLC) can invoke the corresponding script directly. The resulting model package can be handed to the App Builder Skill to generate a local WebUI application, or loaded directly through the Python/C++ APIs of QAI AppBuilder.

\section{Experimental Evaluation}\label{sec:4}

This section reports project deployment, fault diagnosis, repair, and recheck, together with an auxiliary case of an Agent workflow driven by Qwen3.8-27B (\S\ref{sec:4.1}, RQ1--RQ2), followed by performance optimization under accuracy constraints (\S\ref{sec:4.2}, RQ3).

\subsection{Deployment and Agent Workflow Evaluation}\label{sec:4.1}

Each experiment follows the project inputs and deliverables defined in \S\ref{sec:3.3} and runs in an independent workspace without artifacts from earlier runs. The records include the original expected output, target-device execution and output checks, API cost, workflow wall time, and HITL.

\textbf{Result and HITL judgment.} \texttt{O} means that the target-device path, declared numerical or task Gate, and critical delivery capabilities all pass; required cache, preprocessing, and postprocessing paths must also be usable. \texttt{?} means that the QAIRT model and scripts run but lack critical capability, sufficient testing, or metric evidence, for example KV-cache, the required accuracy, or the required QNN path after switching to DLC, and requires further work before delivery. \texttt{X} means that the target path, numerical result, or project contract has a substantive error, for example switching to CPU or GPU execution instead of the required target-NPU path, and requires diagnosis and re-execution.

\texttt{0-\allowbreak{}HITL} means that, after project setup is frozen, the Agent completes a run without human intervention that changes the current decision or execution path. Human intervention is limited to environment recovery or confirmation of the established configuration; it does not provide a repair solution. Supplying task information, modifying code or configuration, authorizing a new operation, or appending an instruction all count as HITL.

\textbf{Test starting-point assumption.} All deployment tests in this paper start from a PyTorch project workspace that is already runnable; missing packages, dependency conflicts, path errors, or an unrunnable PyTorch baseline are treated as unmet preconditions and are not counted in QNN/QAIRT deployment reliability results.

\textbf{Primary test path.} The deployment experiments use the QNN model-library plus Context Binary path: \texttt{qnn-\allowbreak{}*-\allowbreak{}converter} generates the model source and weights, which are then built into the target \texttt{.\allowbreak{}so}/\texttt{.\allowbreak{}dll}, and a Context Binary is prepared. This path more readily exposes deployment issues in local compilation toolchains, I/O layout, and backend configuration. The model optimization in \S\ref{sec:4.2} uses DLC artifacts and the newer unified QAIRT workflow.

\subsubsection{Purposively Covered Model Set}\label{sec:4.1.1}

This paper purposively selects seven PyTorch projects: vision, Real-ESRGAN~\cite{r41} and YOLO-Worldv2~\cite{r42}; speech, MeloTTS~\cite{r43} and Whisper-tiny~\cite{r44}; diffusion, Stable Diffusion v1.5 (SD1.5)~\cite{r45}; and decoder-only LLMs, SmolLM2-135M~\cite{r46} and Qwen3-0.6B~\cite{r47}. These projects involve convolution, multi-branch detection, encoder-decoder structures, autoregressive caches, large UNets, and different I/O interfaces, and are used to examine the AIPC workflow under different deployment conditions. The YOLO-Worldv2 tests keep the CLIP~\cite{r48} component frozen.

Beyond the main experimental models, we performed deployment tests with PPOCR~\cite{r49}, RT-DETR~\cite{r50}, YOLOv5, YOLOv8, and YOLO11~\cite{r51}, YOLO26~\cite{r52}, Wav2Vec2~\cite{r53}, MobileViT~\cite{r54}, LPRNet~\cite{r55}, SmolVLM~\cite{r56}, and Qwen3.5-0.8B~\cite{r57} after freezing the workflow, as supplementary functional checks across different model projects.

\subsubsection{QAIRT Deployment and Failure-Recovery Case Studies}\label{sec:4.1.2}

Tests use an Ubuntu 24.04 LTS x86\_64 host for the Agent, connected over SSH to a QCS9075 (RB8) target device, with QAIRT 2.48.40.260702, Codex CLI~\cite{r58}, and the GPT-5.6 series~\cite{r59}.

This group examines AIPC's target-device execution, output alignment, and feedback-driven diagnosis, repair, and recheck. A vanilla-agent path without the AIPC Skill, but with access to the source project, QAIRT SDK, and official documentation, provides the no-Skill comparison.

In the table, \texttt{Skill=\allowbreak{}Y/\allowbreak{}N} indicates whether the AIPC Skill is used; \texttt{LUNA}, \texttt{TERRA}, and \texttt{SOL} denote \texttt{gpt-\allowbreak{}5.6-\allowbreak{}luna}, \texttt{gpt-\allowbreak{}5.6-\allowbreak{}terra}, and \texttt{gpt-\allowbreak{}5.6-\allowbreak{}sol}, with \texttt{L(low)/\allowbreak{}M(medium)/\allowbreak{}H(high)/\allowbreak{}X(xhigh)} denoting thinking level. \texttt{Result} is \texttt{O/\allowbreak{}?/\allowbreak{}X /\allowbreak{} n;\allowbreak{} H} (pass/pending/fail; planned runs; HITLs); \texttt{Observation} records key symptoms, repair notes, or path deviations; \texttt{?} in \texttt{Time} or \texttt{Cost} denotes that the corresponding data for that run was not completely recorded.

\begin{table*}[!t]
\centering
\footnotesize
\caption{Overview of the deployment process}
\label{tab:4.3}
\begin{tabularx}{\textwidth}{>{\raggedright\arraybackslash\hspace{0pt}\hsize=0.864\hsize}Xcll>{\centering\arraybackslash\hspace{0pt}\hsize=1.136\hsize}X>{\centering\arraybackslash\hspace{0pt}\hsize=0.932\hsize}X>{\raggedright\arraybackslash\hspace{0pt}\hsize=1.068\hsize}X}
\toprule
\textbf{Case} & \textbf{Skill} & \textbf{Result} & \textbf{LLM} & \textbf{Time (min)} & \textbf{Cost (USD)} & \textbf{Observation} \\
\midrule
Real-ESRGAN & Y & \mbox{O3/?0/X0~/~3;~H0} & \texttt{LUNA-\allowbreak{}X} & 19.8/\allowbreak{}18.4/\allowbreak{}17.7 & 1.9/\allowbreak{}1.9/\allowbreak{}1.7 &  \\
Real-ESRGAN & N & \mbox{O2/?0/X1~/~3;~H0} & \texttt{LUNA-\allowbreak{}X} & 18.7/\allowbreak{}7.7/\allowbreak{}8.1 & 1.3/\allowbreak{}0.5/\allowbreak{}0.5 & Unexpected output \\
YOLO-Worldv2 & Y & \mbox{O3/?0/X0~/~3;~H0} & \texttt{LUNA-\allowbreak{}X} & 18.0/\allowbreak{}19.0/\allowbreak{}22.6 & 1.9/\allowbreak{}2.2/\allowbreak{}2.3 &  \\
YOLO-Worldv2 & N & \mbox{O1/?2/X0~/~3;~H0} & \texttt{LUNA-\allowbreak{}X} & 14.0/\allowbreak{}7.0/\allowbreak{}10.2 & 1.0/\allowbreak{}0.5/\allowbreak{}0.7 & Unexpected output: 2; ambiguous configuration: 1 \\
MeloTTS & Y & \mbox{O0/?1/X0~/~1;~H0} & \texttt{TERRA-\allowbreak{}M} & 128.0 & 13.5 & Waveform gate not met \\
MeloTTS & Y & \mbox{O0/?1/X0~/~1;~H1} & \texttt{LUNA-\allowbreak{}X} & 261.0 & 5.858 & HITL: relaxed gate to stop retries \\
MeloTTS & N & \mbox{O0/?1/X1~/~2;~H1} & \texttt{LUNA-\allowbreak{}X} & 32.9 /\allowbreak{} 43.0 & 3.4 /\allowbreak{} ? & VTCM oversized; chunked-decoder repair; waveform gate not met \\
Whisper-tiny & Y & \mbox{O3/?0/X0~/~3;~H0} & \texttt{LUNA-\allowbreak{}X} & 19.1 /\allowbreak{} 26.8 /\allowbreak{} 25.4 & 2.4 /\allowbreak{} 2.2 /\allowbreak{} 2.1 &  \\
Whisper-tiny & N & \mbox{O0/?3/X0~/~3;~H0} & \texttt{LUNA-\allowbreak{}X} & 21.5 /\allowbreak{} 28.7 /\allowbreak{} 18.6 & 0.5 /\allowbreak{} 2.159 /\allowbreak{} 1.003 & No KV-cache export \\
SD1.5 & Y & \mbox{O0/?3/X0~/~3;~H0} & \texttt{LUNA-\allowbreak{}X} & 68.0/\allowbreak{}27.5/\allowbreak{}40.0 & 3.5 /\allowbreak{} 1.5 /\allowbreak{} ? & Cosine similarity <0.99 \\
SD1.5 & N & \mbox{O0/?2/X1~/~3;~H0} & \texttt{LUNA-\allowbreak{}X} & 30.1/\allowbreak{}42.1/\allowbreak{}31.4 & 1.8/\allowbreak{}1.4/\allowbreak{}1.3 & HTP mmap exceeded; workaround: using .so library \\
SmolLM2-135M & Y & \mbox{O2/?0/X0~/~2;~H0} & \texttt{LUNA-\allowbreak{}X} & 38.0/\allowbreak{}117.8 & 1.3/\allowbreak{}7.7 &  \\
SmolLM2-135M & N & \mbox{O0/?2/X1~/~3;~H0} & \texttt{LUNA-\allowbreak{}X} & 37.9 /\allowbreak{} 24.0 /\allowbreak{} 6.3 & 1.8 /\allowbreak{} ? /\allowbreak{} 0.4 & No KV-cache export; Output error: 1 \\
Qwen3-0.6B & Y & \mbox{O1/?0/X0~/~1;~H0} & \texttt{LUNA-\allowbreak{}X} & 99.1 & 2.5 &  \\
Qwen3-0.6B & N & \mbox{O0/?1/X0~/~1;~H0} & \texttt{LUNA-\allowbreak{}X} & 30.0 & 0.8 & Graph split; no KV-cache/Context Binary \\
\bottomrule
\end{tabularx}
\end{table*}

Table~\ref{tab:4.3} summarizes Skill on/off deployment results, process costs, and key observations.

In the experiments, the Skill path can take longer because it includes additional checking steps, whereas no-Skill time is more variable and can increase because of incorrect settings, retries, and repairs. It can also be shortened by switching to the DLC path or skipping AOT. An Agent may also stop after completing a smoke test. YOLO-Worldv2's \texttt{ambiguous configuration: 1} denotes one use of \texttt{SM8550} as the target SoC configuration instead of the physical target \texttt{QCS9075}. SmolLM2-135M used \texttt{qnn-\allowbreak{}pytorch-\allowbreak{}converter}, bypassing the ONNX flow, and produced incorrect results. The WAV generated by MeloTTS is usable, but the waveform Gate did not reach the threshold; the no-Skill path once found a chunked-decoder repair for this model. Several records retained NPY but did not generate expected target deliverables such as images or audio (\texttt{Unexpected output}), or did not cover the KV-cache path. In one of the three SD1.5 no-Skill runs, HTP could not map the complete U-Net model, causing a loading failure; the other two worked around the limit by using the .so model library. Thus, successful model conversion, normal termination of the target process, or the production of intermediate outputs does not mean that the deployment is ready; deployment must also meet the runtime, output-comparison, and deliverable requirements specified by the project.

\begin{table*}[!t]
\centering
\footnotesize
\caption{Representative feedback-driven cases}
\label{tab:4.3b}
\begin{tabularx}{\textwidth}{>{\raggedright\arraybackslash\hspace{0pt}\hsize=0.981\hsize}X>{\raggedright\arraybackslash\hspace{0pt}\hsize=0.981\hsize}X>{\raggedright\arraybackslash\hspace{0pt}\hsize=1.135\hsize}X>{\raggedright\arraybackslash\hspace{0pt}\hsize=0.903\hsize}X}
\toprule
\textbf{Case} & \textbf{Symptom} & \textbf{Diagnosis} & \textbf{Repair} \\
\midrule
YOLO-Worldv2 & Unsupported \texttt{Einsum} & ONNX rewrite & \texttt{Einsum} $\rightarrow$ \texttt{reshape} / \texttt{transpose} / \texttt{matmul} \\
MeloTTS (\texttt{Skill=\allowbreak{}Y}) & VTCM capacity exceeded (\texttt{11.16 MiB > 8 MiB}) & HTP VTCM capacity limit & Convolution time tiling \\
Qwen3-0.6B & Single-graph mapping \texttt{\textasciitilde{}1.50 GiB} & Single-graph size limit & Graph split \\
SD1.5 & U-Net weight mapping fails & HTP/FastRPC weight-mapping limit & Graph split \\
\bottomrule
\end{tabularx}
\end{table*}

Table~\ref{tab:4.3b} summarizes representative feedback-driven paths from symptom to diagnosis and repair.

\textbf{Answer to RQ2.} Table~\ref{tab:4.3b} shows that feedback can trigger the Agent's cross-layer diagnosis, repair, and recheck. The MeloTTS temporal-chunking case corresponds to a \texttt{Skill=\allowbreak{}Y} record in Table~\ref{tab:4.3}: it removes the decoder VTCM limit and restores target HTP execution, but remains \texttt{?} because the waveform Gate does not reach its threshold. For SD1.5, graph splitting removes the full U-Net weight memory-mapping limit. Numerical, task, or application Gates after the repair independently determine final delivery status. A later Gate failure indicates that accuracy or the complete application path still needs work; it does not negate removal of the original deployment blocker.

\subsubsection{Auxiliary Experiment: Feasibility of a Qwen3.8-27B-Driven Agent Workflow}\label{sec:4.1.3}

\textbf{Experiment purpose and case positioning.} This section examines the ability of Qwen3.8-27B, a model capable of local deployment, to drive an AIPC deployment workflow (experiments invoke it through OpenRouter), reporting the end-to-end records of three test models and a vanilla-agent comparison.

\textbf{Execution configuration.} The experiments run on the Snapdragon X Elite Plus Windows platform with Pi Code Agent~\cite{r60} (medium thinking) and QAIRT 2.47. The only difference between the two settings is whether the Agent receives the \textbf{AIPC Skill, including its workflow prompts}; all other conditions are held constant. The tasks are Real-ESRGAN, YOLO-Worldv2, and Whisper-tiny from \S\ref{sec:4.1.2}; each setting is planned for 3 independent runs, and the current Whisper-tiny record contains 3 Skill runs and 1 no-Skill run.

\begin{table*}[!t]
\centering
\footnotesize
\caption{Qwen3.8-27B Agent Workflow Results}
\label{tab:4.4}
\begin{tabularx}{\textwidth}{>{\raggedright\arraybackslash\hspace{0pt}\hsize=0.826\hsize}Xcl>{\centering\arraybackslash\hspace{0pt}\hsize=1.087\hsize}X>{\raggedright\arraybackslash\hspace{0pt}\hsize=1.087\hsize}X}
\toprule
\textbf{Model} & \textbf{Skill} & \textbf{Result} & \textbf{Time (min)} & \textbf{Observation} \\
\midrule
Real-ESRGAN & Y & \mbox{O3/?0/X0~/~3;~H0} & 39.0/\allowbreak{}23.1/\allowbreak{}29.0 &  \\
Real-ESRGAN & N & \mbox{O2/?0/X1~/~3;~H0} & 25.5/\allowbreak{}16.2/\allowbreak{}X69.5 & Python inference code missing \\
YOLO-Worldv2 & Y & \mbox{O3/?0/X0~/~3;~H0} & 40.0/\allowbreak{}38.7/\allowbreak{}68.2 & False incompatibility. \\
YOLO-Worldv2 & N & \mbox{O0/?1/X2~/~3;~H1} & 52.3/\allowbreak{}X28.7/\allowbreak{}X14.6 & Postprocessing ROI error; drifted to DLC/GPU \\
Whisper-tiny & Y & \mbox{O2/?0/X1~/~3;~H0} & 47.8/\allowbreak{}20.0/\allowbreak{}X10.0 & Decoder on CPU. \\
Whisper-tiny & N & \mbox{O0/?1/X0~/~1;~H4} & 110.5 & DLC path; encoder-only stop \\
\bottomrule
\end{tabularx}
\end{table*}

Table~\ref{tab:4.4} contrasts executions of the same Qwen3.8-27B under Skill on/off settings.

An \texttt{X} prefix on a time value denotes a failed run.

We observed three patterns in these experiments. \textbf{(1) Path ambiguity.} When tools were invoked through an x86-emulated Python environment, Qwen3.8-27B sometimes identified the target device as x86 and shifted to CPU-simulated or GPU execution. It could also switch to a DLC path after an error on the QNN path.

\textbf{(2) Anomalous execution time.} \texttt{unsupported} messages from the ONNX version and false alarms triggered additional diagnosis, graph surgery, and validation, lengthening workflow wall time. The X14.6m YOLO-Worldv2 run shifted to DLC/GPU; the X10.0m Whisper-tiny run placed the entire decoder on the CPU, creating an HTP/CPU mixed pipeline that did not meet the target-NPU contract.

\textbf{(3) Delivery and interface-generation failures.} The X69.5m Real-ESRGAN run used DLC/Context Binary through the CLI after multiple repairs but did not generate Python inference code. The no-Skill 110.5m Whisper-tiny run used a DLC path and stopped repeatedly during encoder conversion, decoder conversion, and end-to-end inference; it returned to HTP only after human intervention. The YOLO-Worldv2 postprocessing detection-ROI error also required manual repair.

\subsubsection{Summary}\label{sec:4.1.4}

The two groups of tests show that the AIPC Skill affects delivery by constraining path selection and checks.

\textbf{Without the AIPC Skill (\texttt{Skill=\allowbreak{}N}).} An Agent relying only on QAIRT documentation and general tools can still complete conversion and target-side execution, and even perform graph splitting or interface repair, but may also turn to shortcut paths such as DLC, fixed shapes, no KV-cache, or NPY-only output. These paths sometimes shorten a local step but fail to meet the delivery expectations implied by the project, thus producing \texttt{?} or \texttt{X} and requiring humans to realign the goal.

\textbf{With the AIPC Skill (\texttt{Skill=\allowbreak{}Y}).} AIPC encodes the project contract, deployment steps, tool invocations, and Gates as executable constraints, allowing the Agent to handle graph surgery, runtime artifacts, and target-device checks in one workflow. For autoregressive models, the Skill can require explicit KV-cache export and match the tensor I/O of the target NPU, so each decode step reuses the existing cache rather than recomputing self-attention over the whole prefix. This reduces per-step computation from $O(n^2)$ over the full prefix to $O(n)$ over the cached history.

The Qwen3.8-27B Skill on/off comparison shows the same pattern: \texttt{Skill=\allowbreak{}Y} runs more often remained on the deployment paths required by the project and completed delivery; \texttt{Skill=\allowbreak{}N} runs shifted to DLC/GPU, formed HTP/CPU mixed execution, or omitted Python inference code.

In addition, Qwen3.8-27B can produce executable artifacts and attempt repairs; compared with GPT-5.6-Luna, it currently requires more Harness support for stable execution.

\textbf{Answer to RQ1.} The two groups of cases show that the Agent can deploy the tested PyTorch projects as target-device inference paths. The Skill organizes path selection and checks as executable constraints, allowing more recorded runs to complete delivery along the paths required by the project.

\subsection{QAIRT Accuracy and Performance Optimization}\label{sec:4.2}

This section examines whether AIPC's optimization knowledge and strategies can guide the Agent and translate into accuracy and performance gains (RQ3). The task specification is fixed and each session receives a single prompt; the Skill arm, which can read the Skill, is compared with a control arm that has the same tools and permissions but cannot read it (called Agent-only for YOLO and No-Skill for Whisper). \textbf{EQ-B1}: can the Agent reliably use the recorded knowledge and pass the delivery gates? \textbf{EQ-B2}: can executable gates detect silent failures that are ``loadable and runnable but not deliverable''? \textbf{EQ-B3}: can the triggered optimizations reduce end-to-end latency while satisfying the accuracy and backend constraints?

\begin{table*}[!t]
\centering
\footnotesize
\caption{Experimental design}
\label{tab:4.5}
\begin{tabularx}{\textwidth}{>{\raggedright\arraybackslash\hspace{0pt}\hsize=0.226\hsize}X>{\raggedright\arraybackslash\hspace{0pt}\hsize=1.696\hsize}X>{\raggedright\arraybackslash\hspace{0pt}\hsize=1.077\hsize}X}
\toprule
\textbf{Item} & \textbf{YOLO-Worldv2} & \textbf{Whisper-Small} \\
\midrule
Precision / SDK & W8A8 / QAIRT 2.48.40.260702 & W8A16 / QAIRT 2.51.40.260930 (aligned with AI Hub) \\
Agent model (thinking level) & gpt-5.6-luna (xhigh) & glm-5.3~\cite{r61} (no thinking), deepseek-v4.1-flash~\cite{r62} (high), claude-5-5-sonnet~\cite{r63} (medium) \\
Runs & 14: 4 per arm in the accuracy series; 3 per arm in the performance series, starting from the deliverable of each arm's first accuracy run (fixed before the performance runs began; 96.3\% of FP32 for the Skill arm, 85.8\% for the Agent-only arm), with accuracy criteria judged against this baseline & 13: Skill 7, No-Skill 6 \\
Data & COCO val2017~\cite{r64}: calibration 400 / development 1,000 / final 3,600 images; FP32 mAP@0.5:0.95 = 0.3702 & LibriSpeech~\cite{r65}: final 300 utterances (test-clean); FP32 WER 2.784\% \\
Pre-set criteria & mAP@0.5:0.95 and mAP@0.5 $\geq$ 95\% of FP32, plus per-class recall, whole-class disappearance, and empty-detection-rate criteria; accuracy drop $\leq$ 1.0 percentage point in the performance series & Loadable; WER $\leq$ FP32 + 0.217 percentage points; decoder and encoder each $\leq$ 1.2$\times$ the same-session AI Hub precompiled package \\
Timing & batch 1, NPU backend only, HTP burst, 3 warm-up runs; preprocessing / NPU / postprocessing timed separately & Native I/O; each binary in its own process, measured alternately with the AI Hub package in the same session \\
\bottomrule
\end{tabularx}
\end{table*}

\textbf{Design (Table~\ref{tab:4.5}).} All runs use the same task specification (except the starting artifacts of the two performance-series arms), the Agent built into QAI AppBuilder, thresholds, SDK, and device configuration, and are evaluated only on held-out data that took no part in development or threshold setting. The platform is Snapdragon X2 Elite (Hexagon NPU, HTP V81) with Windows 11 ARM64. We additionally perform three kinds of supplementary builds (one each, not counted in the statistics) for attribution: a naive reference (the original model plus default quantization commands), single-factor ablations, and operator-level profiling.

\subsubsection{Case 1: YOLO-Worldv2}\label{sec:4.2.1}

\begin{table}[!t]
\centering
\small
\caption{YOLO-Worldv2 results (final evaluation set, 3,600 images)}
\label{tab:4.6}
\begin{tabularx}{\columnwidth}{>{\raggedright\arraybackslash\hspace{0pt}\hsize=1.136\hsize}X>{\centering\arraybackslash\hspace{0pt}\hsize=0.932\hsize}X>{\centering\arraybackslash\hspace{0pt}\hsize=0.932\hsize}X}
\toprule
\textbf{} & \textbf{Skill} & \textbf{Agent-only} \\
\midrule
\textbf{(a) Accuracy series (4 runs per arm)} &  &  \\
Per-run mAP@0.5:0.95 (relative to FP32) & 97.2\%, 96.4\%, 87.0\%, 96.5\% & 73.6\%, 76.7\%, 96.4\%, 85.5\% \\
Overall-accuracy criterion ($\geq$ 95\% of FP32) & \textbf{3/4} & \textbf{1/4} \\
All criteria & 1/4 & 1/4 \\
\textbf{(b) Performance series (3 runs per arm)} &  &  \\
End-to-end latency (ms, start $\rightarrow$ end) & 18.16 $\rightarrow$ 6.48 & 13.85 $\rightarrow$ 8.57 \\
Median of per-run speedups (range) & 2.791$\times$ (2.768--2.819) & 1.599$\times$ (1.369--1.692) \\
Accuracy change & 0.00 percentage points & $-$0.16 to $-$0.15 percentage points \\
\bottomrule
\end{tabularx}
\end{table}

Note: the performance series reports medians over 3 runs per arm, each starting from that arm's pre-fixed baseline artifact. All 14 run artifacts are loadable, and metadata confirms that they execute on the NPU. The two arms of the performance series start from different points (18.16 and 13.85 ms), so speedups are compared rather than absolute end-point latencies. In the accuracy series, ``All criteria'' means satisfying A1--A5 together; in the performance series, the criteria are judged against each arm's own starting artifact and require A3--A5, A6, and a consistency check on the confidence threshold, but not A1/A2. The criteria are: A1 mAP@0.5:0.95 $\geq$ 95\% of FP32; A2 mAP@0.5 $\geq$ 95\% of FP32 (MLPerf Inference rule~\cite{r66}); A3 no large drop in recall for any of the 78 classes; A4 no whole class disappears; A5 empty-detection rate $\leq$ FP32 + 0.02; A6 accuracy drop $\leq$ 1.0 percentage point. A1 and A2 together form the ``overall-accuracy criterion'', and A3--A5 prevent the average from masking degradation in individual classes.

\textbf{Accuracy.} Three of the 4 Skill runs satisfy the overall-accuracy criterion, versus 1 in the Agent-only arm (small sample, difference not significant, p = 0.49). With per-class recall and the other criteria added, only 1 run per arm passes all criteria; the passing Skill run confirmed the DLC target strategy at the start, following the Skill workflow (one other run adopted the same strategy without asking but did not pass).

\textbf{Latency.} The Skill arm saves 11.69 ms end to end, of which postprocessing accounts for 76\%, preprocessing 13\%, and the NPU only 11\%; the Agent-only arm saves 5.29 ms. The 2.791$\times$ therefore comes mainly from the host side, not from NPU acceleration.

\begin{table*}[!t]
\centering
\footnotesize
\caption{Main changes triggered by the Skill and their adoption ($^{\dagger}$ = silent failure)}
\label{tab:4.7}
\begin{tabularx}{\textwidth}{>{\raggedright\arraybackslash\hspace{0pt}\hsize=1.314\hsize}X>{\centering\arraybackslash\hspace{0pt}\hsize=0.462\hsize}X>{\raggedright\arraybackslash\hspace{0pt}\hsize=1.225\hsize}X}
\toprule
\textbf{Change} & \textbf{Adoption (Skill vs. control)} & \textbf{Measured effect} \\
\midrule
Boxes and scores sharing one output $\rightarrow$ separate outputs $^{\dagger}$ & 4/4 vs. 4/4 & The naive reference detects no boxes; 74.6\% after splitting \\
Activation calibration min-max $\rightarrow$ entropy & 4/4 vs. 2/4 & +8.4 percentage points \\
Scores output before Sigmoid (logit domain) $^{\dagger}$ & 3/4 vs. 1/4 & Delivered artifacts 96.4\%--97.2\% vs. 73.6\%--87.0\% \\
2 fused outputs $\rightarrow$ 6 multi-scale head outputs $^{\dagger}$ & 3/4 vs. 1/4 & +6.55 percentage points alone in ablation \\
Per-channel weight quantization with CLE $^{\dagger}$ & both arms & +5.95 percentage points with 6 outputs (96.2\%) \\
NCHW $\rightarrow$ NHWC input/output & 3/3 vs. 0/3 & Critical-path cycles $-$7.4\% \\
float32 $\rightarrow$ native uint8 input (LUT) & 3/3 vs. 0/3 & Preprocessing 4.11 $\rightarrow$ 2.61 ms \\
Filter by score before decoding & 3/3 vs. 2/3 & Postprocessing 9.95 $\rightarrow$ 1.05 ms \\
Context compilation target must be read back and confirmed $^{\dagger}$ & 0/7 vs. 3/7 fell back & The fallback artifact is 9\% slower on the NPU \\
\bottomrule
\end{tabularx}
\end{table*}

\textbf{Which knowledge triggered the optimizations (Table~\ref{tab:4.7}).} Both arms split the outputs and filter before decoding; the differences concentrate in changes that depend on NPU quantization and compilation mechanics. Several of these are \textbf{silent failures} (marked $^{\dagger}$): the artifact still loads normally but accuracy collapses or performance degrades. Controlled ablations show that the passing recipe is a conjunction of several changes: without split outputs, accuracy is 0.0\% (not a single box is detected in 3,600 images); the other tested missing-item builds reach 74.6\%--90.2\%, and the best build (96.2\%) approaches the 96.4\%--97.2\% delivered by the Agent.

Note: the first five rows belong to the accuracy series (4 runs per arm); the NHWC, uint8, and filtering rows belong to the performance series (3 runs per arm); the last row spans both series (7 runs per arm). Adoption was verified by reading back the metadata of the delivered artifacts.

\subsubsection{Case 2: Whisper-Small}\label{sec:4.2.2}

\begin{table*}[!t]
\centering
\footnotesize
\caption{Whisper-Small W8A16, 13 runs (final evaluation set, 300 utterances)}
\label{tab:4.8}
\begin{tabularx}{\textwidth}{lccccc>{\centering\arraybackslash\hspace{0pt}\hsize=1.000\hsize}Xc}
\toprule
\textbf{Arm} & \textbf{Loadable} & \textbf{WER met} & \textbf{Decoder met} & \textbf{Encoder met} & \textbf{All criteria} & \textbf{Encoder ms} & \textbf{Decoder ms} \\
\midrule
Skill & 7/7 & 7/7 & 7/7 & 6/7 & \textbf{6/7} & 29.68--36.69 & 2.92--3.63 \\
No-Skill & 4/6 & 3/6 & 1/6 & 0/6 & \textbf{0/6} & 38.63--378.87 & 3.91--42.93 \\
Naive reference & Yes & Yes & Yes & No & No & 63.54 & 3.35 \\
\bottomrule
\end{tabularx}
\end{table*}

Note: the same-session AI Hub precompiled package has encoder 27.32--27.92 ms and decoder 5.33--6.02 ms. The 2 No-Skill runs that delivered no loadable artifact are counted as failed in every criterion column.

Six of the 7 Skill runs satisfy all criteria, versus 0/6 in the No-Skill arm. All Skill runs meet the WER criterion, and the encoder is 1.07--1.08$\times$ the AI Hub package, with the single miss at 1.31$\times$. The No-Skill arm fails mainly on latency: the encoder of its loadable runs is 1.41--13.72$\times$ the AI Hub package, 3 of them slower than the naive reference, and 2 further runs delivered no loadable artifact. The difference between the two arms has a Fisher exact test p = 0.0047. In addition, 300 utterances sampled from LibriSpeech test-clean with no overlap with the original evaluation set were used to re-test all 13 artifacts, again giving 6/7 versus 0/6; across 16 combinations of WER and encoder thresholds, the conclusion is unchanged in 12, and only when the encoder bound is relaxed to 1.5$\times$ does one No-Skill run pass.

\textbf{Latency is mainly explained by one quantization option.} The naive reference has the same WER as FP32 but an encoder 2.32$\times$ the AI Hub package. Changing only the activation quantization schema (\texttt{-\allowbreak{}-\allowbreak{}act\_\allowbreak{}quantizer\_\allowbreak{}schema}) to \texttt{unsignedsymmetric} covers 91\% of the naive-to-expert latency gap (63.54 $\rightarrow$ 33.12 ms). All 7 Skill runs adopted this schema and none of the 6 No-Skill runs did --- exactly the knowledge the Skill records for this model family.

\subsubsection{Discussion: Cost and Limitations}\label{sec:4.2.3}

\textbf{Cost.} The Skill does not bring a consistent efficiency advantage: the active time of the Skill arm is similar to or longer than that of the control arm (Whisper 73.5 vs. 67.6 minutes). But counted per qualifying artifact, the gap is decisive: the 6 Whisper No-Skill runs consumed 920 thousand output tokens and 490 minutes and yielded 0 artifacts; the 7 Skill runs used 875 thousand tokens and 477 minutes and yielded 6, or 146 thousand tokens and 79.5 minutes per artifact.

\textbf{Answer to RQ3.} \textbf{EQ-B1}: when the methods are recorded in the Skill, the Agent uses them reliably. Whisper all-criteria is 6/7 vs. 0/6; for YOLO both arms pass all criteria equally often (1/4 vs. 1/4), and the Skill arm leads only on overall accuracy (3/4 vs. 1/4). \textbf{EQ-B2}: 5 of the 9 changes are silent failures; the artifacts load normally and are identified afterwards by task-level checks, independent evaluation, or metadata readback. \textbf{EQ-B3}: in both cases latency improves while satisfying the accuracy constraint and keeping NPU execution: for Whisper the gain comes from the NPU computation itself (almost entirely explained by one quantization option), whereas for YOLO it comes mainly from the host CPU, with the NPU-side gain coming from the NHWC layout.

\textbf{Limitations.} The benefit of the Skill depends on the model and backend knowledge recorded in it and does not necessarily apply to all models. The sample is small (1--4 runs per cell, one SoC, one SDK version per model), so the conclusions should not be extrapolated. The contributions of the Skill's structure and of model-specific recipes cannot be separated: the experiments have only Skill and no-Skill arms, with no workflow-only arm (Config, Pass, Gate, checks, rollback, and records without model-specific recipes). The results therefore show that recording a correct recipe in the Skill makes the Agent adopt it (most clearly for Whisper, where one quantization option explains most of the gap), not that the Config--Pass--Gate workflow itself is the main source of the gain. A workflow-only ablation is left to future work.

\section{Discussion and Conclusion}\label{sec:5}

\subsection{Engineering Implications and Future Directions}\label{sec:5.1}

AIPC advances the workflow with the evidence produced at each stage. Conversion results, evidence of HTP backend execution, output comparisons, project artifacts, and profiling all serve as inputs to later decisions; Gates map this evidence to states that trigger continuation, termination, or a transition to an appropriate Pass; subsequent Passes rebuild affected artifacts when needed. Runtime errors, numerical deviations, and missing artifacts therefore change the subsequent execution path.

The Agent's exploration is confined within an acceptable execution scope by the AIPC Skill, executable checks, budgets, and HITL conditions. Humans provide domain-specific contracts, exception policies, and knowledge during project setup and later maintenance; under a fixed contract and configuration, some runs show that the evidence-driven workflow can support the Agent executing within the set boundaries without human intervention. Current Agents may still drift off the path, so after a run completes, humans can review the results against the run records, generated artifacts, and Gate states. Under this division of labor, the Agent can help reduce the manual burden and knowledge threshold of specialized-hardware deployment.

\textbf{Engineering lessons.} The measurements in \S\ref{sec:4.2} support four lessons directly usable for similar deployments. First, successful conversion does not mean deliverable, and acceptance must be task-level and blocking: the default YOLO recipe ran all 3,600 images with zero errors yet detected no box at all, and the naive Whisper build matched FP32 accuracy but its encoder was 2.32$\times$ slower than the AI Hub package. Second, gates must be able to abort the flow rather than only record deviations: across the two cases, 5 of 14 Skill-arm runs still delivered artifacts when the audit gave no pass verdict, and the Agent disclosed this itself in every case. Third, knowledge belongs in audited, tested scripts and not only in documents: a single option coupling activation and weight calibration was never tried in the right direction by the Agent across 22 variants in a run that was later invalidated and used to fix the Skill. Fourth, measure the whole pipeline before tuning the NPU: 89\% of YOLO's 2.791$\times$ speedup comes from host-side pre/postprocessing.

Large composite diffusion pipelines are also constrained by host memory and component size. Taking PixArt-$\Sigma$ DiT (XL/2)~\cite{r67} as an example, its T5~\cite{r68} text encoder exceeds 5 GB; in our small-scale exploration, one path completed the DiT part of the deployment but still kept the T5 encoder on the CPU, while the other attempts failed in the repeated conversion stage. Such results vary with the Agent model, configuration, and path selection, indicating that complex models still require a more stable Harness and a more systematic evaluation.

The current AIPC focuses on adaptation, quantization, and graph-level optimization of existing PyTorch projects. Future work will extend to new model backbones such as Mamba and MoE, and to optimization capabilities such as kernel/operator generation, QAT, and pruning; it will also improve the stability of the Harness and evaluate its applicability and engineering overhead on more models.

\subsection{Conclusion}\label{sec:5.2}

AIPC encodes the target-NPU deployment of a complete PyTorch inference project as a workflow composed of Config, Pass, and Gate. Model Builder establishes a deployment baseline on the target device; the Optimizer generates optimized variants from runtime feedback under accuracy, project-interface, and target-backend constraints. The Agent selects Passes and tools according to workflow state.

On Qualcomm QAIRT on-device NPUs, this paper reports target-backend execution, output comparison, fault diagnosis, and profiling results for vision, speech, LLM, and diffusion tasks. Some fixed-contract runs achieve 0-HITL. In two optimization cases, AIPC's optimization knowledge and strategies guide the Agent to accuracy and performance gains (Whisper-Small all criteria 6/7 vs. 0/6; YOLO-Worldv2 speedup 2.791$\times$, mainly from the host side). The Agent auxiliary experiment with Qwen3.8-27B shows that this model can drive the deployment workflow.

These results support the following engineering conclusion: runtime feedback can drive cross-layer diagnosis, candidate repair, and re-checking of the model graph, project interface, or runtime configuration. AIPC thus provides an implementable engineering path for ``AI Deployment by AI''.

\section*{Data Availability}

QAI AppBuilder, in which AIPC's Model Builder workflow is implemented, is publicly released by Qualcomm under the BSD-3-Clause license~\cite{r24}. The Model Optimization capability described in \S\ref{sec:3.2} is an internal, non-open-source extension, so the results of \S\ref{sec:4.2} cannot be fully reproduced by third parties. The supplementary run records referred to in Section~\ref{sec:4} (per-run ledgers, gate reports, logs, numerical comparisons, and the per-model verification-task matrix) are retained by the authors; some of the supplementary material is restricted by internal infrastructure and non-open-source tooling and cannot currently be made public.

\section*{Acknowledgements}

We thank the Multi-Media Computer Vision Team for their support, and also thank former colleagues Ioannis Nousias, Mark Muir, and Adrian Giura; their long-term collaborative working style provided important engineering inspiration for this paper.

Generative AI tools were used to assist with translation and language editing of this manuscript. All technical content, experimental records, and conclusions were produced or reviewed by the authors, who take full responsibility for the content of this paper.


\clearpage
\begin{thebibliography}{99}
\bibitem{r1} John Nickolls, Ian Buck, Michael Garland, and Kevin Skadron. \emph{Scalable Parallel Programming with CUDA}. In \emph{ACM SIGGRAPH 2008 Classes}, pp. 1--14, 2008.
\bibitem{r2} Adam Paszke et al. \emph{PyTorch: An Imperative Style, High-Performance Deep Learning Library}. NeurIPS, 2019.
\bibitem{r3} Leming Shen, Qiang Yang, Xinyu Huang, Zijing Ma, Yuanqing Zheng, and Chris Xiaoxuan Lu. \emph{Automating Model Deployment on Mobile Devices}. In \emph{The 24th Annual International Conference on Mobile Systems, Applications and Services (MobiSys Workshop '26)}, June 21--25, 2026, Cambridge, United Kingdom. ACM, 2026.
\bibitem{r4} Shunyu Yao et al. \emph{ReAct: Synergizing Reasoning and Acting in Language Models}. ICLR, 2023.
\bibitem{r5} Jianhao Su, Zhanwei Wu, Shengting Huang, and Weidong Feng. \emph{AIPC: Agent-Based Automation for AI Model Deployment with Qualcomm AI Runtime}. arXiv:2604.14661, 2026. \url{https://arxiv.org/abs/2604.14661}.
\bibitem{r6} Jianhao Su, Zhanwei Wu, Chia-Heng Tu, and ShengTing Huang. \emph{From PyTorch to the NPU: LLM-Agent-Driven Model Conversion Across Heterogeneous Inference Runtimes}. arXiv:2609.27249v1, 2026. \url{https://arxiv.org/html/2609.27249v1}.
\bibitem{r7} ONNX Project. \emph{ONNX: Open Neural Network Exchange}. \url{https://onnx.ai/}
\bibitem{r8} Microsoft Corporation. \emph{ONNX Runtime: Cross-Platform, High Performance ML Inferencing and Training Accelerator}. \url{https://onnxruntime.ai/}
\bibitem{r9} Mergen Nachin et al. \emph{ExecuTorch - A Unified PyTorch Solution to Run ML Models On-Device}. MLSys 2026.
\bibitem{r10} Google. \emph{LiteRT}. \url{https://developers.google.com/edge/litert}. Accessed 2026-10-13.
\bibitem{r11} Xiaotang Jiang et al. \emph{MNN: A Universal and Efficient Inference Engine}. MLSys, 2020.
\bibitem{r12} Tianqi Chen et al. \emph{TVM: An Automated End-to-End Optimizing Compiler for Deep Learning}. OSDI, 2018.
\bibitem{r13} Chris Lattner et al. \emph{MLIR: A Compiler Infrastructure for the End of Moore's Law}. arXiv:2002.11054, 2020. \url{https://arxiv.org/abs/2002.11054}.
\bibitem{r14} IREE Authors. \emph{IREE: Intermediate Representation Execution Environment}. \url{https://iree.dev/}. Accessed 2026-10-13.
\bibitem{r15} Microsoft. \emph{Olive: Simplify ML Model Finetuning, Conversion, Quantization, and Optimization for CPUs, GPUs and NPUs}. \url{https://github.com/microsoft/Olive}. Accessed 2026-10-05.
\bibitem{r16} NVIDIA Corporation. \emph{NVIDIA TensorRT SDK}. \url{https://developer.nvidia.com/tensorrt}. Accessed 2026-10-24
\bibitem{r17} Qualcomm Technologies Inc. \emph{Qualcomm AI Engine Direct SDK}. \url{https://www.qualcomm.com/developer/software/qualcomm-ai-engine-direct-sdk}. Accessed 2026-10-24
\bibitem{r18} Apple Inc. \emph{Core ML Documentation}. \url{https://developer.apple.com/documentation/coreml}. Accessed 2026-10-13.
\bibitem{r19} Rockchip Electronics. \emph{RKNPU2: Rockchip Neural Processing Unit SDK}. \url{https://github.com/airockchip/rknn-toolkit2/tree/master/rknpu2}. Accessed 2026-10-24
\bibitem{r20} Intel Corporation. \emph{OpenVINO Toolkit: Open Visual Inference and Neural Network Optimization}. \url{https://docs.openvino.ai/}. Accessed 2026-10-24
\bibitem{r21} Huawei Technologies Co., Ltd. \emph{CANN: Compute Architecture for Neural Networks}. \url{https://www.hiascend.com/en/software/cann}. Accessed 2026-10-13.
\bibitem{r22} Sangeetha Siddegowda et al. \emph{Neural Network Quantization with AI Model Efficiency Toolkit (AIMET)}. arXiv:2201.08442, 2022. \url{https://arxiv.org/abs/2201.08442}.
\bibitem{r23} Qualcomm Technologies, Inc. \emph{Qualcomm AI Hub}. \url{https://aihub.qualcomm.com/}. Accessed 2026-10-13.
\bibitem{r24} Qualcomm. \emph{QAI AppBuilder (Quick AI Application Builder)}. GitHub, BSD-3-Clause. \url{https://github.com/qualcomm/qai-appbuilder}. Accessed 2026-10-24.
\bibitem{r25} ggml-org. \emph{llama.cpp}. \url{https://github.com/ggml-org/llama.cpp}. Accessed 2026-10-13.
\bibitem{r26} AMD. \emph{ROCm AI}. \url{https://www.amd.com/en/products/software/rocm/rocm-ai.html}. Accessed 2026-10-10.
\bibitem{r27} AMD. \emph{ROCm Hyperloom: An Agentic System That Auto-Optimizes LLM Workloads on AMD GPUs}. \url{https://github.com/AMD-AGI/Hyperloom}. Accessed 2026-10-10.
\bibitem{r28} Ascend Community. \emph{Agent Skills: AI Agent Toolkit for Huawei Ascend NPU}. \url{https://github.com/Ascend/agent-skills}
\bibitem{r29} Marcin Spoczynski et al. \emph{Xe-Forge: Multi-Stage LLM-Powered Kernel Optimization for Intel GPU}. arXiv:2605.26118v1, 2026.
\bibitem{r30} Jiajie Li, Erwei Wang, Zhiru Zhang, and Samuel Bayliss. \emph{From Human Guidance to Autonomy: Agent Skill System for End-to-End LLM Deployment on Spatial NPUs}. Machine Learning for Architecture and Systems Workshop (MLArchSys), co-located with ISCA 2026. arXiv:2606.07586v2, 2026.
\bibitem{r31} Mengling Deng et al. \emph{EdgeFM: Efficient Edge Inference for Vision-Language Models}. arXiv:2604.27476v2, 2026.
\bibitem{r32} Taras Sereda et al. \emph{KForge: LLM-Driven Cross-Platform Kernel Generation for AI Accelerators}. MLArchSys Workshop at ISCA 2026. arXiv:2606.02963v1, 2026.
\bibitem{r33} Genghan Zhang et al. \emph{AccelOpt: A Self-Improving LLM Agentic System for AI Accelerator Kernel Optimization}. arXiv:2511.15915v2, 2026.
\bibitem{r34} Terry Chen et al. \emph{AVO: Agentic Variation Operators for Autonomous Evolutionary Search}. arXiv:2603.24517v1, 2026.
\bibitem{r35} Alec M. Hammond et al. \emph{Agentic Operator Generation for ML ASICs}. MLSys 2026. arXiv:2512.10977v1, 2025.
\bibitem{r36} Kirill Nagaitsev, Luka Grbcic, Samuel Williams, and Costin Iancu. \emph{Optimizing PyTorch Inference with LLM-Based Multi-Agent Systems}. MLSys 2026. arXiv:2511.16964v2, 2026.
\bibitem{r37} Harsh Menon et al. \emph{Wave: A Symbolic Python DSL And Compiler for High-Performance Machine Learning}. MLSys 2026.
\bibitem{r38} Ignacio Cano et al. \emph{CATWILD: Compiler Autotuning for TPU Workloads in the Wild}. MLSys 2026.
\bibitem{r39} Yuran Ding, Ruobing Han, Xiaofan Zhang, and Xinwei Chen. \emph{PROMPTS: PeRformance Optimization via Multi-Agent Planning for LLM Training and Serving}. MLSys 2026.
\bibitem{r40} Sung Min Cho et al. \emph{Optimizing Deployment Configurations for LLM Inference}. MLSys 2026.
\bibitem{r41} Xintao Wang et al. \emph{Real-ESRGAN: Training Real-World Blind Super-Resolution with Pure Synthetic Data}. ICCVW, 2021.
\bibitem{r42} Tianheng Cheng et al. \emph{YOLO-World: Real-Time Open-Vocabulary Object Detection}. CVPR, 2024.
\bibitem{r43} Wenliang Zhao, Xumin Yu, and Zengyi Qin. \emph{MeloTTS: High-Quality Multi-Lingual Multi-Accent Text-to-Speech}. Software release, 2023. \url{https://github.com/myshell-ai/MeloTTS}.
\bibitem{r44} Alec Radford et al. \emph{Robust Speech Recognition via Large-Scale Weak Supervision}. arXiv:2212.04356, 2022. \url{https://arxiv.org/abs/2212.04356}.
\bibitem{r45} Robin Rombach et al. \emph{High-Resolution Image Synthesis with Latent Diffusion Models}. CVPR, 2022.
\bibitem{r46} Hugging Face. \emph{SmolLM2-135M Model Card}. \url{https://huggingface.co/HuggingFaceTB/SmolLM2-135M}. Accessed 2026-10-13.
\bibitem{r47} An Yang et al. \emph{Qwen3 Technical Report}. arXiv:2505.09388, 2025. \url{https://arxiv.org/abs/2505.09388}.
\bibitem{r48} Alec Radford et al. \emph{Learning Transferable Visual Models From Natural Language Supervision}. ICML, 2021.
\bibitem{r49} PaddlePaddle. \emph{PaddleOCR}. \url{https://github.com/PaddlePaddle/PaddleOCR}. Accessed 2026-10-19.
\bibitem{r50} Yian Zhao, Wenyu Lv, Shangliang Xu, Jinman Wei, Guanzhong Wang, Qingqing Dang, Yi Liu, and Jie Chen. \emph{DETRs Beat YOLOs on Real-time Object Detection}. CVPR, 2024.
\bibitem{r51} Ultralytics. \emph{Ultralytics YOLO Documentation}. \url{https://docs.ultralytics.com/}. Accessed 2026-10-19.
\bibitem{r52} Glenn Jocher, Jing Qiu, Mengyu Liu, Shuai Lyu, Fatih Cagatay Akyon, and Muhammet Esat Kalfaoglu. \emph{Ultralytics YOLO26: Unified Real-Time End-to-End Vision Models}. arXiv:2606.03748, 2026. \url{https://arxiv.org/abs/2606.03748}.
\bibitem{r53} Alexei Baevski, Yuhao Zhou, Abdelrahman Mohamed, and Michael Auli. \emph{wav2vec 2.0: A Framework for Self-Supervised Learning of Speech Representations}. NeurIPS, 2020.
\bibitem{r54} Sachin Mehta and Mohammad Rastegari. \emph{MobileViT: Light-weight, General-purpose, and Mobile-friendly Vision Transformer}. ICLR, 2022.
\bibitem{r55} Sergey Zherzdev and Alexey Gruzdev. \emph{LPRNet: License Plate Recognition via Deep Neural Networks}. arXiv:1806.10447, 2018. \url{https://arxiv.org/abs/1806.10447}.
\bibitem{r56} Hugging Face. \emph{SmolVLM-500M-Instruct Model Card}. \url{https://huggingface.co/HuggingFaceTB/SmolVLM-500M-Instruct}. Accessed 2026-10-19.
\bibitem{r57} Qwen. \emph{Qwen3.5-0.8B Model Card}. \url{https://huggingface.co/Qwen/Qwen3.5-0.8B}. Accessed 2026-10-13.
\bibitem{r58} OpenAI. \emph{Codex CLI: Lightweight Coding Agent in the Terminal}. \url{https://github.com/openai/codex}
\bibitem{r59} OpenAI. \emph{Models}. \url{https://developers.openai.com/api/docs/models}. Accessed 2026-10-13.
\bibitem{r60} earendil-works. \emph{Pi: AI Agent Toolkit and Coding Agent}. \url{https://github.com/earendil-works/pi}. Accessed 2026-10-13.
\bibitem{r61} Z.AI. \emph{GLM-5.3}. \url{https://docs.z.ai/guides/llm/glm-5.3}. Accessed 2026-10-24.
\bibitem{r62} DeepSeek. \emph{DeepSeek-V4.1-Flash-1017 Model Card}. \url{https://huggingface.co/deepseek-ai/DeepSeek-V4.1-Flash-1017}. Accessed 2026-10-24.
\bibitem{r63} Anthropic. \emph{Models overview}. \url{https://platform.claude.com/docs/en/models/overview}. Accessed 2026-10-23.
\bibitem{r64} Tsung-Yi Lin et al. \emph{Microsoft COCO: Common Objects in Context}. ECCV, 2014.
\bibitem{r65} Vassil Panayotov, Guoguo Chen, Daniel Povey, and Sanjeev Khudanpur. \emph{Librispeech: An ASR Corpus Based on Public Domain Audio Books}. ICASSP, 2015.
\bibitem{r66} MLCommons. \emph{MLPerf Inference Rules}. \url{https://github.com/mlcommons/inference_policies/blob/master/inference_rules.adoc}. Accessed 2026-10-24.
\bibitem{r67} Junsong Chen et al. \emph{PixArt-$\Sigma$: Weak-to-Strong Training of Diffusion Transformer for 4K Text-to-Image Generation}. ECCV, 2024.
\bibitem{r68} Colin Raffel et al. \emph{Exploring the Limits of Transfer Learning with a Unified Text-to-Text Transformer}. Journal of Machine Learning Research, 21(140):1--67, 2020.
\end{thebibliography}
\end{document}